\documentclass[11pt,twoside]{article}
\usepackage{amsmath,amsfonts,amssymb}
\usepackage{latexsym}
\usepackage{dcolumn}
\usepackage{graphicx,epsfig}
\usepackage{amsthm}
\usepackage{color}

\begin{document}

\setcounter{page}{1}

\pagestyle{plain} \vspace{1cm}

\begin{center}
{\large{\bf {Thermodynamics of Non-Commutative Scalar-Tensor-Vector Gravity Black holes}}}\\

\vspace{1cm}
{\bf Sara Saghafi$^{a}$}\footnote{s.saghafi@stu.umz.ac.ir},\quad {\bf K. Nozari$^{a,}$}\footnote{knozari@umz.ac.ir},\quad {\bf M. Hajebrahimi$^{a,}$}\footnote{m.hajebrahimi@stu.umz.ac.ir}\\
\vspace{0.5cm}
$^{a}$Department of Physics, Faculty of Basic Sciences, University of Mazandaran,\\
P. O. Box 47416-95447, Babolsar, Iran\\

\end{center}

\vspace{1.5cm}

\begin{abstract}
In this paper we analyze thermodynamic stability of Schwarzschild Modified Gravity (MOG) black holes in a non-commutative framework. We show that, unlike a commutative MOG black hole, in the coherent state picture of non-commutativity MOG black holes are thermodynamically stable. At the final stage of evaporation
a stable remnant with zero temperature and finite entropy is left in this noncommutative framework. Also, we consider Parikh-Wilczek tunneling mechanism of massive particles from non-commutative MOG black holes and demonstrate that information leaks out of non-commutative MOG black holes in the form of some non-thermal correlations.\\

\end{abstract}
\newpage

\section{Introduction}
\label{sec:intro}

In the last years, for understanding several open cosmological issues such as the late time positively accelerated expansion of the
universe ~\cite{1}-~\cite{4}, increasing attention has been paid to modified theories of gravity. Some of those theories modify general relativity by replacing the scalar curvature by a generic function of $R$ as $f(R)$, or functions of the Riemann and Ricci tensors or even their derivatives ~\cite{5}. So it
is quite natural to ask about black hole features in those gravitational theories
since, in one hand, some black hole signatures may specifically result from the Einstein's gravity and others may
be strong characteristics of all generally covariant theories of gravity. From another point of view, the obtained
results may eliminate some modified gravity models which will be in disagreement with expected or known physical
results. For those purposes, research on thermodynamical quantities of black hole is of particular
interest. \\

Scalar-tensor-vector gravity theory ~\cite{6}, also known as Modified Gravity (MOG),not to be confused"
with Bekenstein's TeVeS ~\cite{7}, is another of those modified theories of gravity based on an action principle that postulates the existence of a vector field, while elevating the three constants of the theory to scalar fields. In the weak-field approximation, STVG produces a Yukawa-like modification of the gravitational force due to a point source. STVG has been used successfully to explain galaxy rotation curves~\cite{8}, the mass profiles of galaxy clusters~\cite{8}, gravitational lensing in the Bullet Cluster~\cite{10} and cosmological observations~\cite{11} without the need for dark matter. Solving the field equations for Scalar-Tensor-Vector-Gravity (STVG) or modified gravity (MOG) results in a static, spherically symmetric black hole solution defined by the mass $M$ with two horizons~\cite{11}. As represented in Ref.~\cite{12} the static, spherically symmetric solution describes the final stage of the collapse of a body in terms of an enhanced gravitational constant $G$ and a gravitational repulsive
force with a charge $Q =\sqrt{\alpha G_N M}$, where $\alpha$ is a parameter defined by $G = G_N(1 + \alpha)$, $G_N$ is Newton's
constant and $M$ is the total mass of the black hole. The black hole has two horizons for $\alpha> 0$ and in contrast to Reissner-Nordstrom solution, when $
M<Q_e$ ($Q_e$ is the electric charge) there
is no horizon-free solution with a naked singularity. In Ref.~\cite{13} black hole thermodynamics in modified gravity have been analyzed. To prevent the violation of the second law of thermodynamics, an entropy should be considered for black holes. Also black holes evaporate on account of Hawking radiation and have a corresponding temperature. They have shown that there is no zero-temperature remnant remaining after Hawking evaporation of black hole and the temperature of the black hole diverges at the final stage of evaporation. In consequent, there is no resolution of the information loss paradox in their setup. Finally they have concluded that a modified gravity black hole is thermodynamically unstable. \\

In this paper, we analyze the effect of non-commutativity on the
thermodynamics of a MOG black hole. Theoretical discovery of radiating black holes~\cite{14} has opened a new window on the mysteries of quantum gravity. After several years of intensive research in this field (see Ref.~\cite{15} for a seminal review with an extensive
reference list) various aspects of the problem still remain under question. For example, a full and satisfactory description
of the late stage of black hole evaporation is still missing and needs lots of work to be explored. The string/black hole correspondence principle ~\cite{16} suggests
that in this extreme regime stringy effects cannot be neglected and are actually dominant. There are various outcomes of string
theory, but we focus on the result that the coordinates of spacetime  change to be noncommuting operators on a D-brane ~\cite{17}.
Thus, string-brane coupling has revealed the necessity of spacetime quantization.
This indication resembles the older ideas firstly proposed in a paper by Snyder ~\cite{18}.
The noncommutativ spacetime can be deduced from the following commutator

\begin{equation}\label{comu}
[X^\mu,X^\nu]=\textit{i} \theta^{\mu\nu}\,,
\end{equation}
where $ \theta^{\mu\nu}$ is a matrix which is anti-symmetric and can be determined to discretize the fundamental cell of spacetime much similar to the Planck constant $\hbar$ quantize the phase space. This issue was motivated firstly by the need to control the divergences appearing
in theories such as quantum electrodynamics. In \cite{19} it is shown that $ \theta^{\mu\nu}$ can be assumed as $ \theta^{\mu\nu}=\theta diag(\epsilon_{\imath\jmath},\epsilon_{\imath\jmath}...)$, where
$\theta$ is a constant with dimension
of length squared. It has been demonstrated that noncommutativity
replaces smeared objects with point-like structures in flat spacetime. The
authors have shown in Ref.~\cite{19,20} that the effect of smearing has to be shown in
such a way that a Gaussian
distribution of minimal width $\sqrt{\theta}$ should be considered instead of the position Dirac-delta function everywhere. In this framework, the mass density
of a smeared, particle-like gravitational source which is static and spherically symmetric is written as:

\begin{equation}
\rho_{\theta}(r)=\frac{M}{(2\pi\theta)^{3/2}}\exp(-\frac{r^2}{4\theta})
\end{equation}

Upon integration one obtains the total mass $M$ of matter distribution, and that in the limit $\theta\rightarrow0$ one recovers Dirac's delta function. As has been indicated in Ref.~\cite{19,20}, the particle mass $M$ is spread through a region of linear size $\sqrt{\theta}$ ,instead of being localized at a point. This is
due to the intrinsic uncertainty as has been shown in the coordinate commutators (\ref{comu}). Phenomenological results, so far, imply that noncommutativity is not detectable at presently accessible energies, i.e.
$\sqrt{\theta} < 10^{-16} cm$. Therefor, there are minimal deviations from the standard vacuum Schwarzschild
solution at large distances. But Then, new physics is anticipated at distances near $r\simeq \sqrt{\theta}$ where the existing density of energy
and momentum is non-negligible. This view point on spacetime noncommutativity has stimulated a lot of research programs
in recent years, some of which can be seen in Refs.~\cite{21}-~\cite{40}. In this paper we adopt this
viewpoint to study thermodynamics of non-commutative modified gravity black holes and also reconsideration of the Parikh-Wilczek tunneling mechanism of massive particles emitted from a non-commutative modified gravity back hole. We show that unlike commutative MOG black hole (which has been studied in Ref.~\cite{13}), these black holes in the noncommutative framework have thermodynamical stability. Also the Hawking radiation in this noncommutative case is not purely thermal and there are correlations between emitted modes that may
provide a trace to address (or shedding light on) the information loss paradox.

\section{Thermodynamics of STVG Black holes in Non-Commutative Framework}

The static spherically symmetric Schwarzschild-MOG metric obtained from the corresponding field equations ~\cite{41,42} is given
by

\begin{equation}
ds^{2}=\Big(1-\frac{2GM}{r}+\frac{GQ^{2}}{r^{2}}\Big)dt^{2}-\Big(1-\frac{2GM}{r}+\frac{GQ^{2}}{r^{2}}\Big)^{-1}dr^2-r^2d\Omega^{2}\,,
\end{equation}
where the usual Newtonian constant $G_N$ gets modified to $G=G_N(1+\alpha)$ where $\alpha$ is a free parameter and $Q = \sqrt{\alpha G_N} M$. This metric exhibits two horizons. In Ref.~\cite{13} authors claimed that although algebraically similar to the Reissner-Nordstrom black hole solution, the MOG metric is fundamentally different in the sense that the gravitational charge in MOG is not independent of the mass of black hole, so it is not possible to have a non-zero temperature remnant left after Hawking evaporation. Also they showed that MOG black hole is thermodynamically unstable. \\

In this section, considering the effects of non-commutativity of spacetime on the singularity and temperature of black hole results in some modifications to be illustrated. In the coherent state picture of non-commutativity, the matter section of Einstein field equations would be affected by non-commutativity. In fact, noncommutativity of geometry would change the matter distribution ~\cite{43,44}. It should be noted that noncommutativity is an intrinsic property of manifold, not an imposed property to the structure of geometry which its effect exhibits in changing the matter-energy distribution. Removing the spacetime singularity at $r=0$ is one of the main consequences of the noncommutative geometry. In fact, the energy-momentum density in a gravitational system is affected by noncommutativity of spacetime so that this energy-momentum density results in restricting the curvature of spacetime to construct a regularized geometry. In this regularized geometry, the black hole radius couldn't get smaller than a specified limit and there would no radius smaller than that limit for black hole. As mentioned above the density of a smeared, symmetric and static mass will be of the form:

\begin{equation}\label{ro}
\rho_{\theta}(r)=\frac{M}{(2\pi\theta)^{3/2}}\exp(-\frac{r^2}{4\theta})\,.
\end{equation}

As it is done in [28] we are going to look for a static, spherically symmetric, asymptotically
Schwarzschild-MOG solution of the Einstein equations, with (\ref{ro}) describing the energy density of the system. In order
to completely define the energy-momentum tensor, we rely on the covariant conservation condition $T^{\mu\nu};\nu=0$. For
spherically symmetric metric this equation turns out to be

\begin{equation}
\partial_{r}T^{r}_{ r}=-\frac{1}{2}g^{00}\partial_{r}g_{00}(T^{r}_{ r}-T^{0}_{ 0})-g^{\theta\theta}\partial_{r}g_{\theta\theta}(T^{r}_{ r}-T^{0}_{ 0})
\end{equation}

To preserve Schwarzschild-like property: $g_{00}=-g_{rr}^{-1}$, we require $T^{r}_{r}=T^{0}_{0}=-\rho_{\theta}(r)$. With this requirement the
divergence free equation allows a solution for $T^{\theta}_{\theta}$ which reads

\begin{equation}
T^{\theta}_{\theta}=-\rho_{\theta}(r)-\frac{r}{2}\partial_{r}\rho_{\theta}(r)
\end{equation}

Rather than a massive, structureless point, a source turns out to a self-gravitating, droplet of anisotropic fluid of
density ρθ, radial pressure $p_{r}=-\rho_{\theta}$ and tangential pressure

\begin{equation}
p_{\perp}=-\rho_{\theta}(r)-\frac{r}{2}\partial_{r}\rho_{\theta}(r)
\end{equation}

n physical grounds, a non-vanishing radial pressure is needed to balance the inward gravitational pull, preventing
droplet to collapse into a matter point. This is the basic physical effect on matter caused by spacetime noncommutativity and the origin of all new physics at distance scale of order $\sqrt{\theta}$. By solving Einstein equations with (\ref{ro}) as a matter source and considering the general form of a spherically symmetric metric in MOG, we find the line element:

\begin{eqnarray} \label{metric}
ds^{2}&=&\bigg(1-\frac{2G_{N}(1+\alpha)M_{\theta}}{r}+\frac{\alpha G_{N}^{2}(1+\alpha) M^{2}_{\theta}}{r^{2}}\bigg)dt^{2}\nonumber\\&-
&\bigg(1-\frac{2G_{N}(1+\alpha)M_{\theta}}{r}+\frac{\alpha G_{N}^{2}(1+\alpha) M^{2}_{\theta}}{r^{2}}\bigg)^{-1}dr^2-r^2d\Omega^{2}\,,
\end{eqnarray}
in which $M_{\theta}=\frac{2M}{\sqrt{\pi}}\gamma(\frac{3}{2},\frac{r^{2}}{4\theta})$ substitutes the $M$ in (\ref{metric}). $M_{\theta}$ is derived from the integral of (\ref{ro}) by changing variables, in which $\gamma(\frac{3}{2},\frac{r^{2}}{4\theta})$ is the lower incomplete gamma function and is defined as:

\begin{equation}
\gamma(\frac{3}{2},\frac{r^2}{4\theta})=\int_{0}^{\frac{r^2}{4\theta}} t^{\frac{1}{2}}e^{-t} dt
\end{equation}

The behaviour of the above metric in large values of $r$ is the way that it would be the ordinary MOG black hole as (3). But in small distances ($r\ll\sqrt{\theta}$), the zeroth component of noncommutative MOG metric would be of the following form

\begin{equation}
g_{00}=1-\frac{M(1+\alpha) r^2}{3\sqrt{\pi}\theta^{\frac{3}{2}}}+O(r^4)\,.
\end{equation}

In fact, the metric defined in (\ref{metric}) in the limit of small distances would change into a de Sitter metric which has a regular behaviour at the $r=0$. In other words, the vacuum energy dependent on the oscillations of non-commutative coordinates would result in an effective cosmological constant as $\Lambda_{eff}=\frac{1+\alpha}{\sqrt{\pi}\theta^{\frac{3}{2}}}$. This effective cosmological constant restricts the infinite curvature of spacetime at $r=0$ which is due to noncommutative effects. As it's clear, at very small distances the effect of gravitational charge in metric is not seen anymore.

The noncommutative MOG black hole metric as given by (\ref{metric}) has the following two horizons

\begin{equation}\label{r}
(r_{_H})_{+}=\frac{(1+\alpha)2M\gamma(\frac{3}{2},\frac{r_{H}^{2}}{4\theta})\Big(1+\sqrt{1-\frac{\alpha}{\alpha+1}}\Big)}{\sqrt{\pi}},
\end{equation}

\begin{equation}
(r_{_H})_{-}=\frac{(1+\alpha)2M\gamma(\frac{3}{2},\frac{r_{H}^{2}}{4\theta})\Big(1-\sqrt{1-\frac{\alpha}{\alpha+1}}\Big)}{\sqrt{\pi}}.
\end{equation}

In this paper we set $G_{N}=1$, therefore $G=1+\alpha$. In the limit of $r_{_H}\gg \sqrt{\theta}$, the horizon radius of MOG black hole is obtained. For considering the effect of non-commutativity on MOG black hole mass, we can extract the relation for mass from equation (\ref{r}) as

\begin{equation}
M(r_{_H})=\frac{(r_{_H})_{_+}\sqrt{\pi}}{2\gamma(\frac{3}{2},\frac{r_{H}^{2}}{4\theta})(1+\alpha)\sqrt{1-\frac{\alpha}{\alpha+1}}}\,.
\end{equation}

\begin{figure}
\flushleft\leftskip1em{
\includegraphics[width=.75\textwidth,origin=c,angle=0]{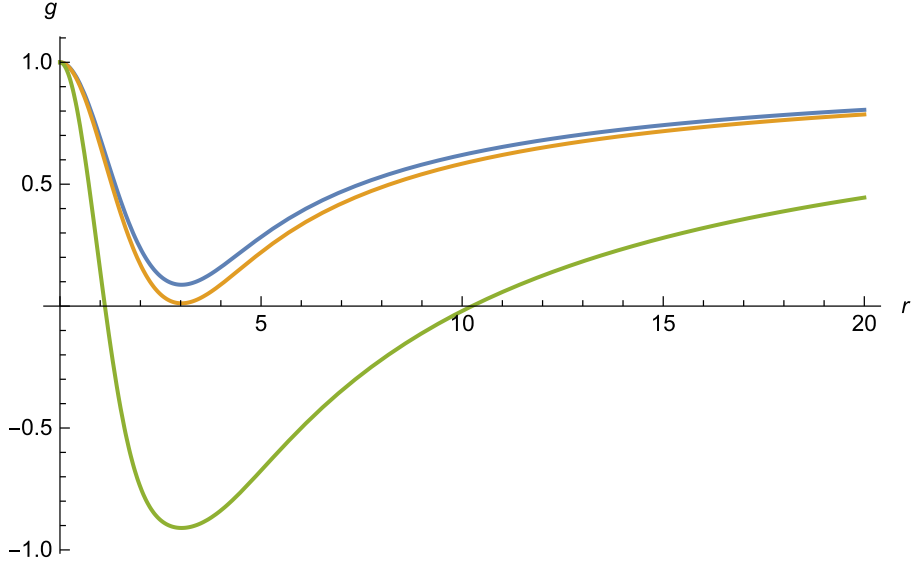}}
\hspace{3.6cm}
\caption{\label{fig1}{\emph{$g_{00}$ versus $r$ for different values of $M$. From the top to the bottom: $M=0.8$ no horizon, $M=1.1$ one degenerate horizon (extremal black hole with horizon $r_{0}$) and $M=3$ two horizons. Note that we have set $\theta=1$.}}}
\end{figure}

 Based on this relation, in the range close to $\sqrt{\theta}$, i.e. where the condition $\frac{r_{_H}}{\sqrt{\theta}}\ll 1$ is fulfilled, there would be a minimum mass $M_0=\frac{1.9\, \theta}{(1+\alpha)\sqrt{1-\frac{\alpha}{\alpha+1}}}$ and a minimum radius $r_0=\frac{3\sqrt{\theta}(1+\alpha)}{2}\Big( 1+\sqrt{1-\frac{\alpha}{\alpha+1}}\Big)$ for non-commutative MOG black hole. For the condition $\alpha=0$, where the Newtonian constant gets no modification and there is no trace of modified gravity, the usual minimum mass for noncommutative Schwarzschild black hole, which is $1.9\,\theta$~\cite{26}-~\cite{40} and the minimum radius $3\sqrt{\theta}$, will be obtained. By considering noncommutativity of spacetime for a MOG black hole, it is obvious that the amount of minimum mass is dependant on two factors; the noncommutativity parameter $\theta$ and the free parameter in MOG, $\alpha$. In Ref.~\cite{13}, it has been claimed that the value of $\alpha$ depends on the strength of gravitational field and the astrophysical object being studied. By setting $\alpha\approx2$ the minimum mass for noncommutative MOG black hole is $M_{0}=1.1\,\theta$ which is smaller than the minimum mass for a noncommutative Schwarzschild black hole. As relation (\ref{r}) does not have any exact analytical solution, for better analyzing the horizon radius of a noncommutative MOG black hole, the plot of $g_{00}$ for different values of $M$ versus $r$ is as shown in the Fig.\ref{fig1}. As this figure demonstrates, the gap between two horizons gets enhanced by increasing $M$. \\

According to Fig.\ref{fig1} the evaporation process of a MOG black hole with the mass $M$ in a noncommutative spacetime can be described as follows:\\
For $M>M_{0}$ the black hole has two horizons, like the ordinary MOG black hole, and therefore it starts to evaporate as usual. For $M=M_{0}$ the evaporation process is over, hence a black hole with a degenerate horizon is left at $r_{0}$. Finally, for $M<M_{0}$ there is no black hole in essence. In other words, an ordinary MOG black hole continues to evaporate towards a zero mass point. But there would be more possibilities for the evaporation process in the noncommutative context for MOG black hole and the evaporation would end after reaching a stable remnant. This remnant could be an address towards resolution of the information loss problem and the unitarity criteria in essence. It means that although the Hawking radiation could be a thermal radiation, it would not last until the zero mass of MOG black hole. Whenever the mass of MOG black hole in the noncommutative spacetime reaches its minimum $M_{0}=1.1\,\theta$, the evaporation process terminates. The information which was gathered during the construction of black hole could be restored in this stable remnant. \\

Let us now consider the black hole temperature by the relation $T_{H}\equiv\frac{1}{4\pi}\frac{dg_{00}}{dr}$ as follows

\begin{eqnarray}\label{T}
T_{H}&=&\frac{1}{4\pi r_{_H}}\bigg(\sqrt{1-\frac{\alpha}{1+\alpha}}\bigg)\bigg(1-\frac{r_{H}^{3}\exp(-\frac{r_{H}^2}{4\theta})}{2\theta^{3/2}\gamma(3/2,\frac{r_{H}^2}{4\theta})}\bigg)\nonumber\\&
&+\frac{\alpha}{4\pi r_{_H}(1+\alpha)\sqrt{1-\frac{\alpha}{1+\alpha}}}
\bigg(-1+\frac{r_{H}^{3}\exp(-\frac{r_{H}^2}{4\theta})}{2\theta^{3/2}\gamma(3/2,\frac{r_{H}^2}{4\theta})}\bigg)
\end{eqnarray}

For large black holes, i.e. $\frac{r_{_H}^2}{4\theta}\gg1$, one recovers the temperature for a MOG black hole which is calculated in Ref.~\cite{13}. Also when $\alpha=0$, the relation (\ref{T}) reduces to the temperature for a noncommutative Schwarzschild black hole. For a noncommutative MOG black hole, the Hawking temperature is dependent on two factors, both the noncommutativity parameter, $\theta$, and the free parameter of the MOG, $\alpha$. \\

\begin{figure}
\flushleft\leftskip1em{
\includegraphics[width=.75\textwidth,origin=c,angle=0]{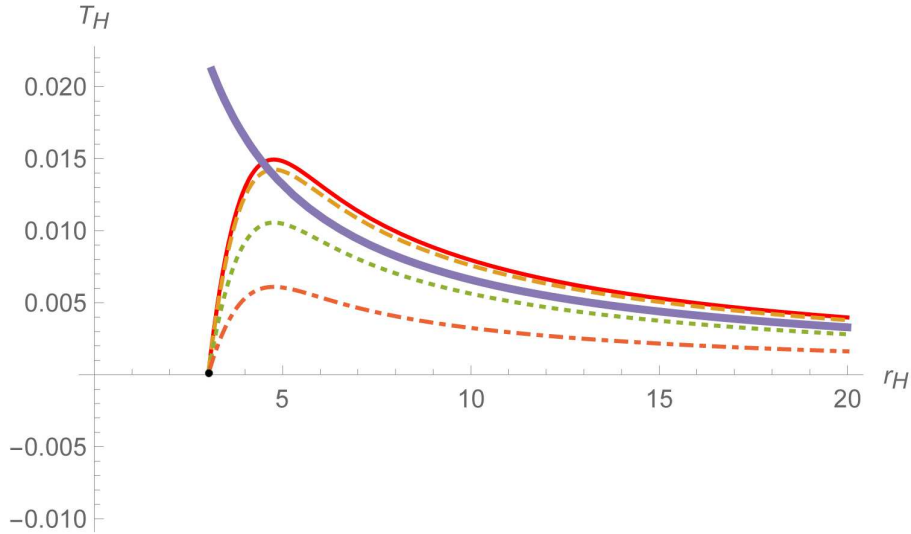}}
\hspace{3.6cm}
\caption{\label{plotT}{ \emph{$T_{H}$ versus $r_{H}$ for different values of the MOG parameter $\alpha$ in a noncommutative framework (the noncommutative parameter $\theta$ is set to $1$). From the top to bottom: the red curve for $\alpha=0$, the yellow dashed curve for $\alpha=0.1$, the green dotted curve for $\alpha=1.98$ and the red dashed-dotted curve for $\alpha=5$. By increasing $\alpha$, the maximum temperature for a noncommutative MOG black hole decreases. For example, for $\alpha=1.98$ the maximum temperature $T_{H}=0.011$ corresponds to the radius $r_{H}=4.73\theta$. For all curves $T_{H}=0$ for $r_{0}\simeq2.97\,\theta$. The purple thick curve is Hawking temperature for a \emph{commutative} MOG black hole.}}}
\end{figure}

At the initial stage of evaporation, the MOG black hole temperature increases while the horizon radius is decreasing. It is
crucial to investigate that what happens as $r_{_H}\rightarrow\sqrt{\theta}$. In the standard commutative case $T_H$ diverges~\cite{13} in the limit of $r\rightarrow 0$ and this is in agreement with the arbitrary description of Hawking radiation. Contrary to this case, temperature (\ref{T})
derived by considering the effects of noncommutative spacetime which are appropriate to distances analogous to $\sqrt{\theta}$. Behavior of the temperature $T_H$
as a function of the horizon radius of MOG black hole is plotted in Fig.\ref{plotT}.

As the Fig.\ref{plotT} shows, the modified Hawking temperature of a noncommutative MOG black hole tends to zero for a stable relic after reaching a maximum temperature. It is demonstrated that $T_{H}=0$ could be obtained for a remnant with minimum horizon radius $r_{0}=2.97\,\theta$ (this is the minimum radius which was obtained firstly in a noncommutative background). The corresponding minimal mass is due to consideration of a Gaussian distribution in a noncommutative background. In the limit $r<r_{0}$ there is no black hole, so that no temperature is defined for this region. Another crucial argument is that by increasing the free parameter $\alpha$ in a MOG black hole, the maximum temperature of a non-commutative MOG black hole is decreasing. In other words, by increasing the gravitational charge, the final state maximum temperature decreases.\\

For calculating the entropy, relation (\ref{T}) should be substituted in the first law of thermodynamics as $dM=T_{H}dS$. But the corresponding integral has not an analytical solution. For calculating the entropy approximately, we consider the standard Hawking temperature $T_{H}=\frac{1}{4\pi r_{_H}}$ but we use the noncommutative MOG Schwarzschild radius $r_{_H}$ as given by (\ref{r}). Another approximation that we applied, is substituting $2M$ in the right hand side of (\ref{r}) instead of $r_{H}$ in the gamma function, for simplicity of integration we use expansion of $\gamma(\frac{3/2}{x})$ according to the error function. We find,
\begin{equation}
T_{H}=\frac{1}{16\sqrt{\pi}M(1+\alpha)\Big(\sqrt{1-\frac{\alpha}{1+\alpha}}\Big)\Big[\frac{\sqrt{\pi}}{2}Erf(\sqrt{\frac{M^2}{\theta}})-\sqrt{\frac{M^2}{\theta}}
\exp(\frac{-M^2}{\theta})\Big]}\,,
\end{equation}
where $Erf(x)$ is the error function. Now the entropy for a non-commutative MOG black hole can be written as
\begin{equation}
S=\frac{12e^{\frac{-M^2}{\theta}}\sqrt{\pi}\sqrt{\frac{M^2}{\theta}}+\Big(4\pi M^2-6\pi\theta\Big)Erf(\sqrt{\frac{M^2}{\theta}})}{\sqrt{\frac{1}{1+\alpha}}}
\end{equation}

\begin{figure}
\flushleft\leftskip1em{
\includegraphics[width=.75\textwidth,origin=c,angle=0]{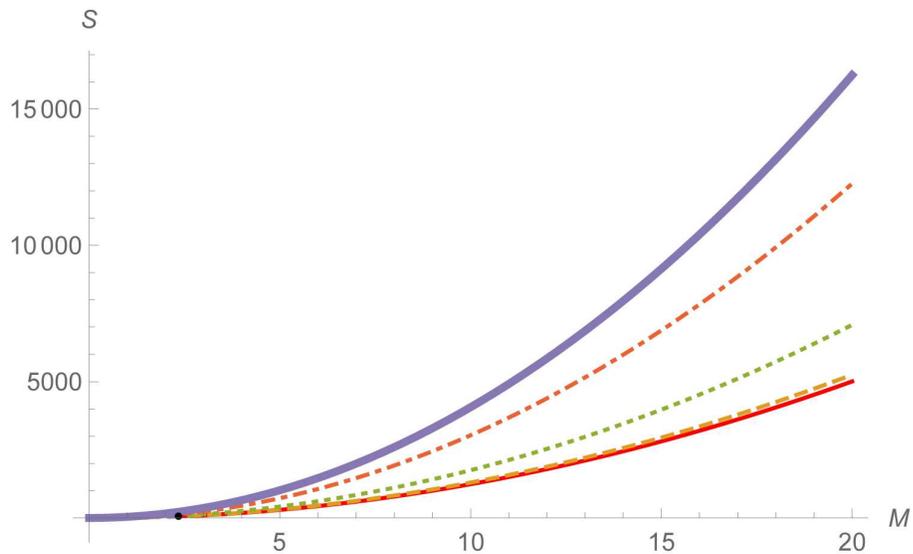}}
\hspace{3.6cm}
\caption{\label{plotS}{\emph{Entropy versus $M$ for different values of $\alpha$ in a noncommutative MOG framework (the noncommutative parameter $\theta$ is set to $1$).
From top to the bottom: the dashed-dotted red curve for $\alpha=5$, the green dotted one for $\alpha=1.98$, the yellow dashed one for $\alpha=0.1$ and the red one for $\alpha=0$. The blue thick curve is entropy for a commutative MOG black hole.}}}
\end{figure}

In the limit of $\frac{M}{\sqrt{\theta}}\gg 1$ the above relation recovers the MOG black hole entropy~\cite{13} without the effects of noncommutativity. Fig.\ref{plotS} exhibits the modified entropy for MOG black hole in the noncommutative framework for different values of MOG parameter $\alpha$. According to this figure, the zero entropy in the last stage of evaporation accompanying a minimum remnant, demonstrates the modification of noncommutativity on a MOG black hole. In fact, the zero entropy means there is one accessible state in which in standard MOG black hole it couldn't be justified because of the total evaporation and reaching to a zero mass for black hole (the blue thick curve in Fig.\ref{plotS}). The modification of spacetime structure in a noncommutative framework easily solves the problem by providing a minimum remnant.\\

A further serach into the role of noncommutativity can be made by looking at the black hole heat capacity $C \equiv \frac{dM}{dT_{H}}$, that can be written as
\begin{equation}
C_{(r_{_H})_{+}}=T_{H}\Big(\frac{dS}{d(r_{_H})_{+}}\Big)\Big(\frac{dT_{H}}{d(r_{_H})_{+}}\Big)^{-1}
\end{equation}

This relation illustrates the black hole exchange of energy with the environment.
This fact that the black hole remnant is not able to exchange
energy with the environment and is an thermodynamically stable object, is the result of any zero root for the temperature corresponds to $C = 0$. On the other hand, any zero root of the relation $\frac{dT_H}{d(r_{_H})_{+}}$  causes to create a pole
and spots a change of sign of the heat capacity, switching the black hole to a $C>0$
stable system, happening a phase transition~\cite{45,46}.

\begin{figure}
\flushleft\leftskip1em{
\includegraphics[width=.45\textwidth,origin=c,angle=0]{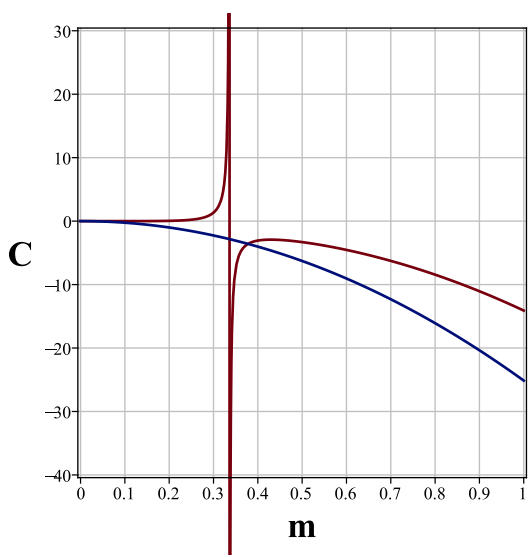}}
\hspace{3.6cm}
\caption {\label{plotC}{\emph{Plot of $C$ versus $r_{H}$ for non-commutative MOG black hole. The vertical line corresponds to a phase transition.}}}
\end{figure}

We have plotted in Fig.\ref{plotC} the specific heat versus the black hole
mass and figured out that the specific heat of a noncommutative MOG black
hole diverges at a point at which the black hole temperature reaches its maximum value and then it decreases to
zero when the mass of the black hole reaches its minimal value. At this point, specific heat vanishes and
thereby the noncommutative MOG black hole does not exchange heat with the surrounding medium leaving a remnant.\\

The modified Hawking temperature $T_H$ in a noncommutative framework can be used in
the calculation of the emission rate. The emission rate might
be calculated using the Stefan-Botlzmann law if the energy
loss is dominated by photons. Assuming a $D$-dimensional
brane embedded in a higher dimensional spacetime, the thermal emission in the bulk can be neglected and the black
hole is supposed to radiate mainly on the brane. It is evident
that in extra dimensional scenarios the final stage of evaporation (black hole remnant) has event horizon area more
than its four dimensional counterpart. Therefore, higher
dimensional black hole remnants have less classical features relative to their four dimensional counterparts~\cite{47,48}. To
obtain the relation between emission rate of black holes
radiation and spacetime dimensionality, we proceed as follows. As Emparan \emph{et al.} have shown in Refs.~\cite{47,48}, in $D$ dimensions, the energy
radiated by a black body of temperature $T$ and surface area $A$ is given by

\begin{equation}
\frac{dE_{D}}{dt}=\sigma_{D}A_{D}T^{D},
\end{equation}

where $\sigma_{D}$ is the $D$-dimensional Stefan-Boltzmann constant
defined as $\sigma_{D}=\frac{\Omega_{D-3}\Gamma(D)\zeta(D)}{(D-2)(2\pi)^{D-1}}$. Therefore, $A=\Omega_{D-2}r_{_H}^{D-2}$ and by definition
$\Omega_{D-2}$ is the metric of the unit $S^{D-2}$ as
$\Omega_{D-2}=\frac{2\pi^{\frac{D-1}{2}}}{\Gamma(\frac{D-1}{2})}$. Now using Eq. (\ref{T}) for modified Hawking
temperature of a non-commutative MOG black hole, it is easy to show
that

\begin{equation}
\frac{dE_{D}}{dt}=\frac{\Omega_{D-3}\Omega_{D-2}}{(2\pi)^{D-1}(D-2)} \Gamma(D)\zeta(D)r_{_H}^{D-2}T^{D}
\end{equation}

For $D =4$, the emission rate can be found as
follows:

\begin{equation}
\frac{dE}{dt}= 4\pi (r_{_H})_{+}^2 T^4\,,
\end{equation}
where can be rewritten as

\begin{eqnarray}
\frac{dM}{dt}&=&4 \pi \bigg(\frac{(1+\alpha)2M\gamma(\frac{3}{2},\frac{r_{H}^{2}}{4\theta})\Big(1+\sqrt{1-\frac{\alpha}{\alpha+1}}\Big)}{\sqrt{\pi}}\bigg)^2\times\nonumber \\ &\bigg(&\frac{1}{16\sqrt{\pi}M(1+\alpha)\Big(\sqrt{1-\frac{\alpha}{1+\alpha}}\Big)\Big[\frac{\sqrt{\pi}}{2}Erf(\sqrt{\frac{M^2}{\theta}})-\sqrt{\frac{M^2}{\theta}}
\exp(\frac{-M^2}{\theta})\Big]}\bigg)^4
\end{eqnarray}

\begin{figure}
\flushleft\leftskip1em{
\includegraphics[width=.75\textwidth,origin=c,angle=0]{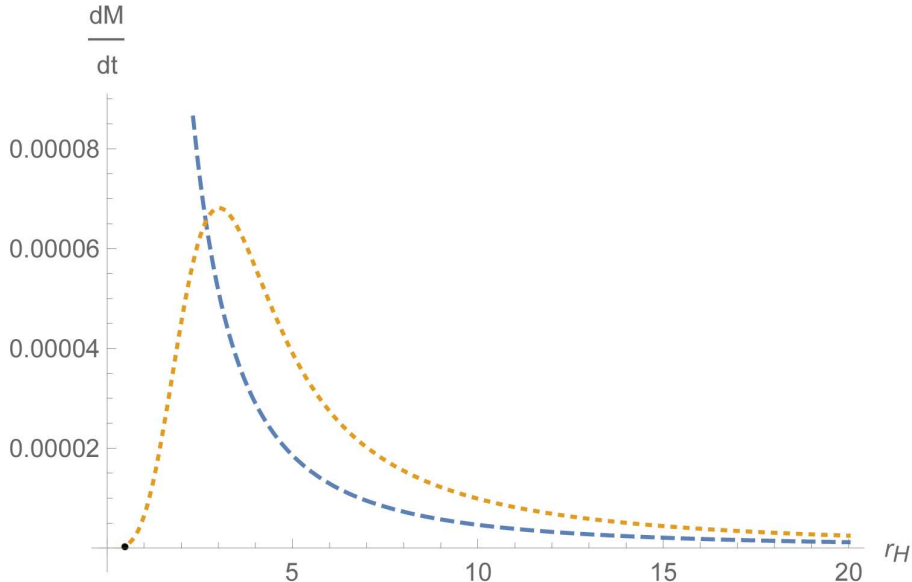}}
\hspace{3.6cm}
\caption{ \label{radiate}{\emph{Plot of emission rate versus $r_{H}$ for non-commutative MOG black hole. The blue dashed one is for commutative case.}}}
\end{figure}

The emission rate of a noncommutative MOG black hole is shown in Fig.\ref{radiate}. This figure shows that the emission rate of a noncommutative MOG black hole vanishes when
the black hole reaches its minimal size. In the standard
framework, the emission rate goes to infinity as the mass of
the commutative MOG black hole tends to zero. In the non-commutative MOG picture, the modified emission rate of black hole doesn't diverge at all, and it
just goes to zero when the black hole's mass reaches its minimal value which can be called as a black hole remnant.

\section{Parikh-Wilchek tunneling mechanism for a MOG black hole in a noncommutative framework}

After the discovery of Hawking radiation~\cite{49}, a lot of attempts have been made to explore
different aspects of this revolutionary achievement. There are some important questions
in this regard: Is black hole radiation exactly thermal? Are unitary and Lorentz invariance
symmetries retained at the quantum gravity level? What happens in the last stage
of black hole evaporation? Are the information that were entered horizon at the time of
black hole formation lost? Although, there is no perfect theory by now to answer these
questions properly, various arguments are presented to address such questions in recent
years. One of these attempts is the strategy provided by Parikh and Wilczek~\cite{50}. In this
approach, particle and antiparticle pairs are created and then particle tunnels through the event
horizon. Due to emission of this particle, total energy of black hole reduces. Conservation of
energy requires that the event horizon radius reduces too. Parikh and Wilczek considered the WKB approximation
to investigate the black hole event horizon thermodynamics. This
approximation is exactly accurate due to an infinite blue shift in the vicinity of the
horizon. Actually, the emitted
particle itself induced the barrier through which tunneling occurs. They have assumed the tunneling particle as a spherically symmetric shell
that is ejected from black hole surface. This approach was the basis of a lot of research
programs then after. Tunneling of massless~\cite{51,52} and massive ~\cite{53} particles from Schwarzschild
black hole and also noncommutative inspired Schwarzschild black hole~\cite{54,55} were studied.
Recently, tunneling of massive and charged particles from Reissner-Nordstr\"{o}m black hole
horizon was reported too~\cite{56}. Also extensions to higher dimensional spacetime models are
considered by some authors~\cite{57}-~\cite{59}. \\

In this section the tunneling of massive particles from the horizon of a noncommutative MOG black hole is studied.
To describe the tunneling process, we require a metric that is not singular on the horizon. The Painlev\'{e} coordinates transformation is a suitable tool to
overcome this difficulty. With the Painlev\'{e} transformation~\cite{60}, the metric (\ref{metric}) takes the following form

\begin{equation}
ds^2=-g_{00}dt_{p}^2+dr^2+2\sqrt{1-g_{00}}dt_{p}dr+r^2d\Omega^{2},
\end{equation}
where $t_{p}$ is the Painlev\'{e} time coordinate and $$g_{00}=\Big(1-\frac{2G_{N}(1+\alpha)M_{\theta}}{r}+\frac{\alpha G_{N}^{2}(1+\alpha) M^{2}_{\theta}}{r^{2}}\Big).$$
With this transformation, horizon's singularity is removed and we can analyze tunneling
process of particles through the horizon. The Lagrangian of a particle with
mass $m$ is written as follows

\begin{equation}
\mathcal{L}=\frac{m}{2}\Big(-g_{00}\dot{t}_{p}^{2}+\dot{r}^{2}+2\sqrt{1-g_{00}}\dot{t}_{p}\dot{r}\Big),
\end{equation}
where a dot marks a the derivative with respect to the proper time, $\tau$.\\
By using the Euler-Lagrange equation ($\frac{\partial
\mathcal{L}}{\partial q}-\frac{d}{dt}\frac{\partial
\mathcal{L}}{\partial\dot{q}}=0$) we find

\begin{equation}\label{eq1}
mg_{00}\dot{t}_{p}-m\sqrt{1-g_{00}}\dot{r}\equiv{\cal{E}}=constant.
\end{equation}

To achieve the radial equation of motion in the Painlev\'{e}
coordinates, we need timelike trajectories that are given by
\begin{equation}\label{eq2}
g_{00}\dot{t}_{p}^{2}+\dot{r}^{2}+2\sqrt{1-g_{00}}\dot{t}_{p}\dot{r}=-1
\end{equation}

Equations (\ref{eq1}) and (\ref{eq2}) can be solved simultaneously to obtain
$\dot{r}$ and $\dot{t}$. Then equation of motion of a massive  particle in Painlev\'{e} coordinates is obtained as follows

\begin{equation}\label{eq3}
\frac{dr}{dt_{p}}=\pm
g_{00}\frac{\sqrt{{\cal{E}}^{2}-m^{2}g_{00}}}{{\cal{E}}\pm\sqrt{1-g_{00}}\sqrt{{\cal{E}}^{2}-m^{2}g_{00}}}.
\end{equation}

As has been mentioned above, there is a infinite blue shift in the vicinity of black hole event
horizon. So, we can use the WKB approximation and calculate coefficients of transmission
for a massive particle that tunnels from inner radius to the outer radius. Firstly,
we consider imaginary part of the action for a particle that is coming from an initial state $r_{in}$ to a final state $r_{out}$

$$\text{Im}\:S\equiv \text{Im}\int
E\:dt=\text{Im}\int_{r_{in}}^{r_{out}}
p_{r}\:dr=\text{Im}\int_{r_{in}}^{r_{out}}\int_0^{p_{r}}\:dp_r\:dr\,.$$

Using the Hamilton equation of motion, $dp_r=\frac{dH}{\dot{r}}$, we find
\begin{equation}\label{integral}
\text{Im}\:S=\text{Im}\int_{r_{in}}^{r_{out}}\int_m^{\cal{E}}\frac{dH}{\dot{r}}\:dr=
-\text{Im}\int_m^{\cal{E}}\int_{r_{in}}^{r_{out}}\frac{dr}{\dot{r}}\:d\tilde{{\cal{E}}}\,,
\end{equation}
where $\tilde{{\cal{E}}}$ is an integration dummy variable. Since the emitted particles are assumed to be massive, the lower
limit of integral now is $m$ instead of being $0$ as for massless
particles. Indeed, before tunneling, spherical
shell of particle has an energy ${\cal{E}}$ and after crossing the
event horizon, changes its energy to ${\cal{E}}-m$ and we will
treat the effect of $m$ on the tunneling rate. We probed motion of a massive particle in Painlev\'{e}'s
coordinates and achieved equation of motion as given by Eq. (\ref{eq3}). Now we expand the metric around the event horizon as
$g_{00}(r)=g_{00}((r_{_H})_{+})+g'_{00}((r_{_H})_{+})(r-(r_{_H})_{+})+...,$ where a prime marks a derivative with respect to the radial coordinate, $r$. Then
Eq. (\ref{eq3}) takes the following form
\begin{equation}\label{eq4}
\dot{r}=\pm g'((r_{_H})_{+})[r-(r_{_H})_{+}]\frac{\sqrt{{\cal{E}}^{2}-m^{2}g'((r_{_H})_{+})[r-(r_{_H})_{+}]}}{{\cal{E}}\pm\sqrt{1-g'((r_{_H})_{+})[r-(r_{_H})_{+}]}\sqrt{{\cal{E}}^{2}
-m^{2}g'((r_{_H})_{+})[r-(r_{_H})_{+}]}}\,.
\end{equation}

By substituting (\ref{eq4}) into Eq. (\ref{integral}), integral has a pole at
$r=r_{_H}$. We solve this integral by using the calculus of residues to find

$$\text{Im} S=\pi\int_m^{\cal{E}}
\frac{2}{g'_{00}((r_{_H})_{+})}\:d\tilde{{\cal{E}}}\,.$$

We note also that, $r_{in}$ and $r_{out}$ are obtained as follows

$$r_{in}=\frac{(1+\alpha)2M \Big[\frac{\sqrt{\pi}}{2}Erf\Big(\sqrt{\frac{M^2}{\theta}}\Big)-\sqrt{\frac{M^2}{\theta}}
\exp\Big(\frac{-M^2}{\theta}\Big)\Big]\Big(1+\sqrt{1-\frac{\alpha}{\alpha+1}}\Big)}{\sqrt{\pi}}$$

$$r_{out}=\frac{(1+\alpha)2(M-{\cal{E}})\Big[\frac{\sqrt{\pi}}{2}\,Erf\Big(\sqrt{\frac{(M-{\cal{E}})^2}{\theta}}\Big)-\sqrt{\frac{(M-{\cal{E}})^2}{\theta}}
\exp\Big(\frac{-(M-{\cal{E}})^2}{\theta}\Big)\Big]\Big(1+\sqrt{1-\frac{\alpha}{\alpha+1}}\Big)}{\sqrt{\pi}}\,.$$

By considering\, $g_{00}=\frac{[r-(r_{_H})_{+}][r-(r_{_H})_{-}]}{r^2}$,\,
the imaginary part of the action takes the following form
\begin{equation}\label{int}
\text{Im}\:S=\pi\int_{m}^{{\cal{E}}}\frac{2
(r_{_H})_{+}}{(r_{_H})_{+}-(r_{_H})_{-}}\:d\tilde{{\cal{E}}}\,.
\end{equation}
Here $(r_{_H})_{+}$ and $(r_{_H})_{-}$ are considered as

$$(r_{H})_{\pm}=\frac{(1+\alpha)2(M-\tilde{\cal{E}})\Big[\frac{\sqrt{\pi}}{2}\,Erf\Big(\sqrt{\frac{(M-\tilde{\cal{E}})^2}{\theta}}\Big)-\sqrt{\frac{(M-\tilde{\cal{E}})^2}{\theta}}
\exp\Big(\frac{-(M-\tilde{\cal{E}})^2}{\theta}\Big)\Big]\Big(1\pm\sqrt{1-\frac{\alpha}{\alpha+1}}\Big)}{\sqrt{\pi}}\,.$$

By Substituting the above relations into the integral (\ref{int}), we obtain the imaginary part
of the action as follows
\begin{eqnarray}
\text{Im}\:S&=&\frac{1}{\sqrt{\frac{1}{1+\alpha}}+\sqrt{\frac{1-2\alpha}{1-\alpha}}} (1+\alpha) \Big(1+\sqrt{\frac{1}{1+\alpha}}\Big)^2\times\nonumber \\
&\Big\{&-\pi\Big[(M-{\cal{E}})-\frac{3\theta}{2}\Big]\, Erf\Big(\frac{M-{\cal{E}}}{\sqrt{\theta}}\Big)-3\sqrt{\pi\theta}(M-{\cal{E}})\exp\Big(-\frac{(M-{\cal{E}})^2}{\theta}\Big)\nonumber \\
&+&\pi\Big[(M-m)-\frac{3\theta}{2}\Big]\, Erf\Big(\frac{M-m}{\sqrt{\theta}}\Big)-3\sqrt{\pi\theta}(M-m)\exp\Big(-\frac{(M-m)^2}{\theta}\Big)\Big\}\,.
\end{eqnarray}

In the WKB approximation, the tunneling probability for the classically prohibited area as a function
of the imaginary part of the particle action at the stationary
phase takes the form $\Gamma\sim\exp(-2\,\text{Im}\:S)$. Therefore, the tunneling rate is

\begin{eqnarray}
\Gamma&\sim& \exp \Bigg\{\frac{-2}{\sqrt{\frac{1}{1+\alpha}}+\sqrt{\frac{1-2\alpha}{1-\alpha}}} (1+\alpha) \Big(1+\sqrt{\frac{1}{1+\alpha}}\Big)^2\times\nonumber \\
&\bigg\{&-\pi\Big[(M-{\cal{E}})-\frac{3\theta}{2}\Big]\, Erf\Big(\frac{M-{\cal{E}}}{\sqrt{\theta}}\Big)-3\sqrt{\pi\theta}(M-{\cal{E}})\exp\Big(-\frac{(M-{\cal{E}})^2}{\theta}\Big)\nonumber \\
&+&\pi\Big[(M-m)-\frac{3\theta}{2}\Big]\, Erf\Big(\frac{M-m}{\sqrt{\theta}}\Big)-3\sqrt{\pi\theta}(M-m)\exp\Big(-\frac{(M-m)^2}{\theta}\Big)\bigg\}\Bigg\}\,.
\end{eqnarray}

The tunneling rate for a non-commutative MOG black hole depends on two parameters $\alpha$ and $\theta$. In case $\alpha=0$ the tunneling rate is for non-commutative Schwarzschild black hole as derived in Ref.~\cite{54,55}. Obviously at high energy regime the emission spectrum of noncommutative MOG black hole
cannot be exactly thermal since the high energy modifications flow from the physics of energy conservation with noncommutativity corrections. In fact the emission rate deviates from a pure thermal emission, however it is corresponding yet to an underlying unitary quantum theory~\cite{61,62}. At this stage, we want to demonstrate that there are some correlations between emitted massive particles with the inclusion of the noncommutativity corrections. This means it can be exhibited that the probability of tunneling of two particles
of energies ${\cal{E}}_{1}$ and ${\cal{E}}_{2}$ and corresponding masses $m_{1}$ and $m_{2}$ is not similar to the probability of tunneling of one particle
with their compound energies, ${\cal{E}}={\cal{E}}_{1}+{\cal{E}}_{2}$ and compound mass $m_{1}+m_{2}$, that is

$$\chi({\cal{E}}_{1}+{\cal{E}}_{2};{\cal{E}}_{1},{\cal{E}}_{2})=\Gamma_{{\cal{E}}_{1}+{\cal{E}}_{2}}-(\Gamma_{{\cal{E}}_{1}}+\Gamma_{{\cal{E}}_{2}})\,.$$

Therefore, the probabilities of emission for a non-commutative MOG black hole are correlated. On the
other hand, the statistical correlation function, $\chi({\cal{E}}; {\cal{E}}_{1}, {\cal{E}}_{2})$ is non-zero which leads to the
dependence between different modes of radiation during the evaporation. Hence, in this
method the form of the corrections of MOG black hole for massive particles with inclusion of noncommutativity effects are adequate by themselves to retrieve information because there
are correlations between different modes and information
comes out with the Hawking radiation. It should be noted that in case $m_{1}=m_{2}=0$, the correlation function becomes zero and there is no correlation between the emitted modes of a noncommutative MOG black hole for massless particles.

\section{Conclusion}
Within the coherent state picture of spacetime noncommutativity, we have studied the thermodynamics and also the Parikh-Wilczek tunneling mechanism of massive particles in a non-commutative MOG black hole. We have shown that noncommutative MOG black hole radiates through Hawking process until it reaches a remnant whose final mass depends on the noncommutativity parameter and the free parameter $\alpha$ in STVG formalism~\cite{13}.  Also we see that within noncommutative geometry the MOG black
hole cools down to a zero temperature remnant after reaching to a maximum temperature and the entropy
vanishes for the remnant mass. The heat capacity and emission rate of MOG black hole in a noncommutative spacetime has been calculated in this paper. As these two quantities can determine the thermodynamical stability of a noncommutative MOG black hole, we figured out that the modified specific heat of black
hole diverges at a point (a phase transition) at which the black hole temperature reaches its maximum value and then it decreases to
zero when the mass of the black hole reaches its minimal value (a remnant). At this point, specific heat vanishes and
thereby the black hole do not exchange heat with the surrounding space leaving a stable remnant. on the other hand the emission rate of the non-commutative black hole vanishes when
the black hole reaches its minimal value. Such a relic could provide potentially a sort of solution to the long standing problem of the
information paradox. Indeed, due to occurrence of phase transition (the cool phase in
place of the hot Planck phase), the initial information is conserved and preserved into
the black hole relic, even after the end of the evaporation. We came to conclusion that the noncommutativity effect becomes more
effective for small values of free parameter $\alpha$, even for a small MOG black hole gravitational charge.
we have also studied tunneling of massive particles in noncommutative horizon of a MOG black
hole. We applied the standard Parikh-Wilczek tunneling method by adopting the Painlev\'{e}
transformation in order to remove singularities of the metric. We have shown that tunneling
rate is dependent on the energy and mass of tunneling particles and also both the noncommutativity parameter $\theta$ and free parameter of MOG $\alpha$ and couldn't be purely thermal anymore. Noncommutativity and tunneling of massive particles could result in correlations between emitted modes for a MOG black hole and the information leaks out of MOG black hole in the form of correlation function. In this paper by considering the tunneling of massive particles of MOG black hole in a noncommutative background, not only the noncommutativity results in stable remnant which could preserve information, but also the tunneling of massive particles of MOG black hole causes to recover some information leaking out of MOG black hole through Hawking radiation. These could be potentially another step towards a complete resolution of the information loss paradox.\\


\begin{thebibliography}{99}

\bibitem{1}
Carroll S M et al, \emph{Phys. Rev. D}  {\bf{70}} (2004) 043528

\bibitem{2} Davli G, Gababdadze G and Porrati M, \emph{Phys. Lett. B} {\bf{485}} (2000) 208

\bibitem{3} de la Cruz-Dombriz A and Dobado A,  \emph{Phys. Rev. D},  {\bf{74}}  (2006) 087501

\bibitem{4}Cembranos J A R,  \emph{Phys. Rev. D}, {\bf{73}}   (2006) 064029

\bibitem{5}
Dobado A and Maroto A L,  \emph{Phys. Lett. B} {\bf316} (1993) 250

\bibitem{6}
J. W. Moffat, \emph{JCAP},  {\bf0603}  (2006) 004, arXiv:0506021 [gr-qc]

\bibitem{7}
J. D. Bekenstein,  \emph{Phys. Rev. D} {\bf70},  (2004) 083509

\bibitem{8}
Brownstein, J. R.; Moffat, J. W,  \emph{Galaxy Rotation Curves Without Non-Baryonic Dark Matter}, \emph{Astrophysical Journal}, {\bf636}  (2006) 721–741,  arXiv:astro-ph/0506370

\bibitem{9}
Brownstein, J. R., Moffat, J. W, \emph{Galaxy Cluster Masses Without Non-Baryonic Dark Matter},  \emph{Monthly Notices of the Royal Astronomical Society}. {\bf367} (2006) 527–540, arXiv:astro-ph/0507222

\bibitem{10}
 Brownstein, J. R., Moffat, J. W, \emph{The Bullet Cluster 1E0657-558 evidence shows Modified Gravity in the absence of Dark Matter}, \emph{Monthly Notices of the Royal Astronomical Society}, {\bf382} (2007) 29–47, arXiv:astro-ph/0702146

\bibitem{11}
Moffat, J. W., Toth, V. T, \emph{Modified Gravity: Cosmology without dark matter or Einstein's cosmological constant}, (2007) ,arXiv:0710.0364

\bibitem{12}
Moffat, J. W, \emph{European Physical Journal C}, {\bf75} (2015) 175, 	arXiv:1412.5424 [gr-qc]

\bibitem{13}
Mureika R. M, Moffat J. w, \emph{Phys. Lett. B}, {\bf757} (2016) 528, 	arXiv:1504.08226 [gr-qc]

\bibitem{14}
S. W. Hawking, \emph{Comm. Math. Phys},  {\bf43} (1975) 199

\bibitem{15}
 T. Padmanabhan, \emph{Phys. Rep},  {\bf406} (2005) 49


\bibitem{16}
L. Susskind, \emph{Phys. Rev. Lett}, {\bf71} (1993) 2367


\bibitem{17}
 E. Witten, \emph{Nucl. Phys. B}, {\bf460} (1996) 335 ; N. Seiberg, E. Witten, \emph{JHEP} {\bf9909} (1999) 032



 \bibitem{18}
H. S. Snyder, \emph{Phys. Rev}, {\bf71} (1947) 38


\bibitem{19}
A. Smailagic and E. Spallucci, \emph{J. Phys. A},  {\bf37} (2004) 1; A. Smailagic and E. Spallucci, \emph{J. Phys. A} {\bf37} (2004) 7169



\bibitem{20} A. Smailagic and E. Spallucci, \emph{J. Phys. A} {\bf36} (2003) L467; A. Smailagic and E. Spallucci, \emph{J. Phys. A} {\bf36} (2003) L517


\bibitem{21}
K. Nozari and S. H. Mehdipour, \emph{Class. Quantum Grav} {\bf25} (2008) 175015

\bibitem{22} Y. G. Miao, Z. Xue and S. J. Zhang, \emph{Gen. Relativ. Gravit} {\bf44} (2012) 555


\bibitem{23}
T. G. Rizzo, \emph{JHEP} {\bf09} (2006) 021

\bibitem{24} K. Nozari and S. H. Mehdipour, \emph{JHEP} {\bf03} (2009) 061

\bibitem{25} K. Nozari and S. H. Mehdipour, \emph{Commun. Theor. Phys} (Beijing, China), {\bf53} (2010) 503

\bibitem{26}
P. Nicolini, A. Smailagic and E. Spallucci, \emph{ESA Spec. Publ}, {\bf637} (2006) 11

\bibitem{27} P. Nicolini, \emph{J. Phys. A},  {\bf38} (2005) L631

\bibitem{28} P. Nicolini, A. Smailagic and E. Spallucci, \emph{Phys. Lett. B}, {\bf632} (2006) 547

\bibitem{29} S. Ansoldi, P. Nicolini, A. Smailagic, \emph{E. Spallucci, Phys. Lett. B}, {\bf645} (2007) 261

\bibitem{30} E. Spallucci, A. Smailagic and P. Nicolini, \emph{Phys. Lett. B}, {\bf670} (2009) 449

\bibitem{31} Y. S. Myung and M. Yoon, \emph{Eur. Phys. J. C},  {\bf62} (2009) 405

\bibitem{32} M. I. Park, \emph{Phys. Rev. D}, {\bf80} (2009) 084026

\bibitem{33} R. Garattini and F. S. N. Lobo, \emph{Phys. Lett. B}, {\bf671} (2009) 146

\bibitem{34} P. Nicolini and E. Spallucci, \emph{Class. Quant. Grav}, {\bf27} (2010) 015010

\bibitem{35} I. Arraut, D. Batic and M. Nowakowski, \emph{Class. Quant. Grav}, {\bf26} (2009) 245006

\bibitem{36} P. Nicolini, \emph{Int. J. Mod. Phys. A}, {\bf24}  (2009) 1229

\bibitem{37} I. Arraut, D. Batic and M. Nowakowski, \emph{J. Math. Phys}, {\bf51} (2010) 022503

\bibitem{38} D. Batic and P. Nicolini, \emph{Phys. Lett. B}, {\bf692} (2010) 32

\bibitem{39} A. Smailagic and E. Spallucci, \emph{Phys. Lett. B}, {\bf688} (2010) 82

\bibitem{40} W. H. Huang, arXiv:1003.1040

\bibitem{41}J. W. Moffat, \emph{Eur. Phys. J. C}, {\bf75} (2015) 175


\bibitem{42} J. W. Moffat, \emph{Eur. Phys. J. C}, {\bf75} (2015) 130



\bibitem{43} J. Michell, \emph{Physical Transaction of the Royal Secety}, {\bf74} (1787) 35


\bibitem{44} P. S. Laplace, \emph{Le System du monde}, {\bf Vol.II} Paris (1795)


\bibitem{45}D. Pavon, \emph{Phys. Rev. D}, {\bf43} (1991) 2495

\bibitem{46} Y. S. Myung, \emph{Mod. Phys. Lett. A}, {\bf23} (2008)  667

\bibitem{47}
Emparan, R., Horowitz, G.T., Myers, R.C., \emph{Phys. Rev. Lett}, {\bf85} (2000) 499, arXiv:hep-th/0003118

\bibitem{48}
Nozari, K., Mehdipour, S. , \emph{Int. J. Mod. Phys} {\bf A 21} (2006) 4979


\bibitem{49}
S. W. Hawking, \emph{Comm. Math. Phys}, {\bf43} (1974) 199


\bibitem{50}
M. K. Parikh and F. Wilczek, \emph{Phys. Rev. Lett}, {\bf85} (2000) 5042


\bibitem{51}
M. Arzano, A. J. M. Medved and E. C. Vagenas, \emph{JHEP} {\bf0509} (2005) 037

\bibitem{52} K. Nozari and S. H. Mehdipour, \emph{Europhys. Lett}, {\bf84} (2008) 20008



\bibitem{53}
J. Zhang and Z. Zhao, \emph{Nucl. Phys. B}, {\bf725} (2005)  173



\bibitem{54}
K. Nozari and S. H. Mehdipour, \emph{Class. Quantum Grav}, {\bf25} (2008) 175015

\bibitem{55} Y. -G. Miao, Z. Xue and S. -J. Zhang, \emph{Gen. Relativ. Gravit.}, {\bf44} (2012) 555



\bibitem{56}
Y. -G. Miao, Z. Xue, S. -J. Zhang, \emph{Europhys. Lett.}, {\bf96} (2011) 10008



\bibitem{57}
T. G. Rizzo, \emph{JHEP}, {\bf09} (2006) 021

\bibitem{58} K. Nozari and S. H. Mehdipour, \emph{JHEP}, {\bf03} (2009) 061

\bibitem{59} K. Nozari and S. H. Mehdipour, \emph{Commun. Theor. Phys.},  {\bf53} (2010) 503



\bibitem{60}
P.Painlev\'{e}, \emph{C. R. Acad. Sci.}, (Paris) {\bf173} (1921) 677



\bibitem{61}
M. K. Parikh, \emph{A Secret Tunnel Through The Horizon}, \emph{Int. J. Mod. Phys. D}, {\bf13} (2004) 2351 ,arXiv:hep-th/0405160



\bibitem{62}
M. K. Parikh, \emph{Energy Conservation and Hawking Radiation}, hep-th/0402166.








\end{thebibliography}
\end{document}